\documentclass[11pt]{article}
\usepackage{amsmath,amssymb,amsthm,amsxtra,overpic,bbm,bm,epsfig,ulem,color,multirow}
\usepackage{braket}

\begin{document}

\begin{center}
{\Large Low scale leptogenesis under neutrino $\mu$-$\tau$ reflection symmetry}
\end{center}

\vspace{0.05cm}

\begin{center}
{\bf Yan Shao, \bf Zhen-hua Zhao\footnote{zhaozhenhua@lnnu.edu.cn}} \\
{ $^1$ Department of Physics, Liaoning Normal University, Dalian 116029, China \\
$^2$ Center for Theoretical and Experimental High Energy Physics, \\ Liaoning Normal University, Dalian 116029, China }
\end{center}

\vspace{0.2cm}

\begin{abstract}
In the literature, the neutrino $\mu$-$\tau$ reflection symmetry (which has the interesting predictions $\theta^{}_{23} =\pi/4$ and $\delta = \pm \pi/2$ for the atmospherical neutrino mixing angle and Dirac CP phase) is an attractive and widely studied candidate for the flavor symmetries in the neutrino sector. But it is known that, when the seesaw model is furnished with this symmetry, the leptogenesis mechanism (which provides an elegant explanation for the baryon-antibaryon asymmetry of the Universe) can only work in the two-flavor regime (which only holds for the right-handed neutrino masses in the range $10^9-10^{12}$ GeV). This prohibits us to have a low scale seesaw model (which has the potential to be directly accessed by running or upcoming collider experiments) that can have the $\mu$-$\tau$ reflection symmetry and successful leptogenesis simultaneously. In this paper, for the first time, we demonstrate that the successful leptogenesis may also be achieved in low scale seesaw models furnished with the $\mu$-$\tau$ reflection symmetry, by means of the flavor non-universality of the conversion efficiencies from the flavored lepton asymmetries to the baryon asymmetry via the sphaleron process. We perform the study in both the resonant leptogenesis regime and the leptogenesis via oscillations (ARS leptogenesis) regime.
\end{abstract}

\newpage

\section{Introduction}

As we know, the non-zero masses of neutrinos along with their flavor mixing are observed in the phenomenon of neutrino oscillations \cite{xing}, which contradicts with the Standard Model (SM) where neutrinos are assumed to be massless. In order to explain the origin of neutrino masses, various theoretical frameworks beyond the Standard Model have been proposed, one of the most popular and natural ways is the type-I seesaw model in which at least two heavy right-handed neutrinos $N^{}_I$ (for $I=1, 2, ...$) are added to the SM \cite{seesaw1}-\cite{seesaw5}.
In this case, the Yukawa interactions between the right- and left-handed neutrinos will lead to the Dirac neutrino mass matrix $M^{}_{\rm D}= Y^{}_{\nu} v$ where $Y^{}_{\nu}$ is the neutrino Yukawa coupling matrix and $v = 174$ GeV is the vacuum expectation value of the Higgs field. Additionally, the right-handed neutrinos have the Majorana mass matrix $M^{}_{\rm R}$ of themselves.
Under the seesaw condition $M^{}_{\rm R} \gg M^{}_{\rm D}$, the effective mass matrix for light neutrinos is obtained as
\begin{eqnarray}
M^{}_{\nu} = - M^{}_{\rm D} M^{-1}_{\rm R} M^{T}_{\rm D} \;.
\label{1}
\end{eqnarray}
Throughout this paper, without loss of generality, we will work in the basis of $M^{}_{\rm R}$ being diagonal as $D^{}_{\rm R} = {\rm diag}(M^{}_1, M^{}_2, M^{}_3)$ with the right-handed neutrino masses in the order $M^{}_1 < M^{}_2 < M^{}_3$.

Diagonalization of Eq.~(\ref{1}) by means of the PMNS mixing matrix $U$ \cite{pmns1, pmns2} gives the light neutrino masses:
\begin{eqnarray}
U^\dagger M^{}_\nu U^* = D^{}_\nu= {\rm diag}(m^{}_1, m^{}_2, m^{}_3) \;,
\label{2}
\end{eqnarray}
with $m^{}_i$ (for $i=1, 2, 3$) being three light neutrino masses. In the standard parametrization, $U$ depends on three mixing angles $\theta^{}_{ij}$ (for $ij=12, 13, 23$), one Dirac CP phase $\delta$ and two Majorana CP phases $\rho$ and $\sigma$ as
\begin{eqnarray}
U  = \left( \begin{matrix}
c^{}_{12} c^{}_{13} & s^{}_{12} c^{}_{13} & s^{}_{13} e^{-{\rm i} \delta} \cr
-s^{}_{12} c^{}_{23} - c^{}_{12} s^{}_{23} s^{}_{13} e^{{\rm i} \delta}
& c^{}_{12} c^{}_{23} - s^{}_{12} s^{}_{23} s^{}_{13} e^{{\rm i} \delta}  & s^{}_{23} c^{}_{13} \cr
s^{}_{12} s^{}_{23} - c^{}_{12} c^{}_{23} s^{}_{13} e^{{\rm i} \delta}
& -c^{}_{12} s^{}_{23} - s^{}_{12} c^{}_{23} s^{}_{13} e^{{\rm i} \delta} & c^{}_{23}c^{}_{13}
\end{matrix} \right) \left( \begin{matrix}
e^{{\rm i}\rho} &  & \cr
& e^{{\rm i}\sigma}  & \cr
&  & 1
\end{matrix} \right) \;,
\label{3}
\end{eqnarray}
where the abbreviations $c^{}_{ij} = \cos \theta^{}_{ij}$ and $s^{}_{ij} = \sin \theta^{}_{ij}$ have been employed. These mixing parameters (excluding the Majorana CP phases) are measured through neutrino oscillation experiments, which are also sensitive to the independent neutrino mass-squared differences: $\Delta m^2_{21}$ and $\Delta m^2_{31}$, where $\Delta m^2_{ij} \equiv m^2_i - m^2_j$ has been defined. The global analysis of neutrino oscillation date is available in Refs.~\cite{global1}-\cite{global3}, and we will use the results in Ref.~\cite{global1} (shown in Table~1 here) as reference values in the following numerical calculations.
The existing neutrino data, as is well known, do not allow to determine the sign of $\Delta m^2_{31}$, thereby allowing for two possible neutrino mass orderings: the normal ordering (NO) $m^{}_1 < m^{}_2 < m^{}_3$ and inverted ordering (IO) $m^{}_3 < m^{}_1 < m^{}_2$.
Furthermore, neutrino oscillation experiments cannot measure the absolute neutrino mass scale and Majorana CP phases. The values of these parameters must be inferred from non-oscillatory experiments.

The particular values of neutrino mixing angles (such as the proximity of $\theta^{}_{23}$ to $\pi/4$) and the Dirac CP phase $\delta \sim 3\pi/2$ \cite{T2K} imply the possible existence of an underlying flavor symmetry in the lepton sector \cite{FS1}-\cite{FS4}. This hints at a deeper structure or organization within neutrino physics, potentially governed by symmetries that could explain these specific values and their implications for neutrino oscillations and CP violating phenomena. One attractive and widely studied candidate of them is the neutrino $\mu$-$\tau$ reflection symmetry \cite{mu-tauR1}-\cite{mu-tauR3}.
Under this symmetry, the neutrino mass matrices keep invariant with respect to the following transformations of three left-handed neutrino fields
\begin{eqnarray}
\nu^{}_{e} \leftrightarrow \nu^{c}_e \;, \hspace{1cm} \nu^{}_{\mu} \leftrightarrow \nu^{c}_{\tau} \;,
\hspace{1cm} \nu^{}_{\tau} \leftrightarrow \nu^{c}_{\mu} \;,
\label{4}
\end{eqnarray}
where the symbol $c$ denotes the charge conjugation of relevant fields.
Such a symmetry leads to the following interesting predictions for the neutrino mixing parameters
\begin{eqnarray}
\theta^{}_{23} = \frac{\pi}{4} \;, \hspace{1cm} \delta = \frac{\pi}{2}  \ {\rm or} \ \frac{3\pi}{2} \;,
\hspace{1cm} \rho = 0 \ {\rm or} \ \frac{\pi}{2} \;, \hspace{1cm} \sigma = 0 \ {\rm or} \ \frac{\pi}{2} \;.
\label{5}
\end{eqnarray}
In the following numerical calculations, we will choose to work with $\delta  = 3\pi/2$ which is more favored experimentally.

On the other hand, via the leptogenesis mechanism \cite{leptogenesis}-\cite{Lreview4}, the extension of SM with the type-I seesaw model can explain the baryon-antibaryon asymmetry of the Universe \cite{planck}
\begin{eqnarray}
Y^{}_{\rm B} \equiv \frac{n^{}_{\rm B}-n^{}_{\rm \bar B}}{s} \simeq (8.69 \pm 0.04) \times 10^{-11}  \;,
\label{6}
\end{eqnarray}
where $n^{}_{\rm B}$ ($n^{}_{\rm \bar B}$) denotes the baryon (antibaryon) number density and $s$ the entropy density. In the seesaw model, the neutrino Yukawa couplings provide the necessary new source of CP violation. The departure from thermal equilibrium will occur when the rates of these Yukawa interactions are slow enough (compared to the expansion rate of the Universe). Lepton number conservation is violated by the Majorana mass terms of these new particles. And the generated lepton asymmetry is converted to a baryon asymmetry via the sphaleron processes.
When the right-handed neutrino mass spectrum is hierarchial, the observed value of $Y^{}_{\rm B}$ can be successfully reproduced for right-handed neutrino mass scale $\gtrsim 10^9$ GeV \cite{DI}, which implies that this scenario cannot be directly accessed by forseeable experiments. Fortunately, it is possible to achieve the successful leptogenesis for TeV scale right-handed neutrinos via the resonant leptogenesis mechanism \cite{resonant1, resonant2}. Furthermore, Akhmedov, Rubakov and Smirnov (ARS) proposed the mechanism of leptogenesis via oscillations \cite{ARS1, ARS2}, in which the observed baryon asymmetry can be generated prior to the electroweak phase transition from the dynamics of GeV scale right-handed neutrinos. In these two regimes, a low scale seesaw model (which has the potential to be directly accessed by running or upcoming collider experiments \cite{RHN}) can accommodate a viable leptogenesis.

However, it is known that \cite{MN}, when the seesaw model is furnished with the above-mentioned $\mu$-$\tau$ reflection symmetry, the leptogenesis mechanism can only work in the two-flavor regime which holds for the right-handed neutrino masses in the range $10^9-10^{12}$ GeV (for detailed explanations, see section~3). This prohibits us to have a low scale seesaw model that can have the $\mu$-$\tau$ reflection symmetry and successful leptogenesis simultaneously. In this paper, we will demonstrate that, by means of the flavor non-universality of the conversion efficiencies from the flavored lepton asymmetries to the baryon asymmetry via the sphaleron process (for detailed explanations, see section~3) \cite{sphaleron1}-\cite{sphaleron4}, the successful leptogenesis may also be achieved in low scale seesaw models furnished with the $\mu$-$\tau$ reflection symmetry.

The remaining parts of this paper are organized as follows. In the next section, we recapitulate some basic formulas of leptogenesis as a basis of our study. In section~3, we first explain why it was thought that under the $\mu$-$\tau$ reflection symmetry the leptogenesis mechanism can only work in the two-flavor regime, and then introduce the flavor non-universality of the conversion efficiencies from the flavored lepton asymmetries to the baryon asymmetry which may help us achieve the successful leptogenesis in low scale seesaw models furnished with the $\mu$-$\tau$ reflection symmetry. In sections~4 and 5, respectively in the resonant and ARS leptogenesis regimes, we study whether the successful leptogenesis can be achieved in low scale seesaw models furnished with the $\mu$-$\tau$ reflection symmetry, with the help of such a property of the sphaleron process. Finally, the summary of our main results will be given in section~6.

\begin{table}\centering
  \begin{footnotesize}
    \begin{tabular}{c|cc|cc}
     \hline\hline
      & \multicolumn{2}{c|}{Normal Ordering}
      & \multicolumn{2}{c}{Inverted Ordering }
      \\
      \cline{2-5}
      & bf $\pm 1\sigma$ & $3\sigma$ range
      & bf $\pm 1\sigma$ & $3\sigma$ range
      \\
      \cline{1-5}
      \rule{0pt}{4mm}\ignorespaces
       $\sin^2\theta^{}_{12}$
      & $0.303_{-0.012}^{+0.012}$ & $0.270 \to 0.341$
      & $0.303_{-0.012}^{+0.012}$ & $0.270 \to 0.341$
      \\[1mm]
       $\sin^2\theta^{}_{23}$
      & $0.451_{-0.016}^{+0.019}$ & $0.408 \to 0.603$
      & $0.569_{-0.021}^{+0.016}$ & $0.412 \to 0.613$
      \\[1mm]
       $\sin^2\theta^{}_{13}$
      & $0.02225_{-0.00059}^{+0.00056}$ & $0.02052 \to 0.02398$
      & $0.02223_{-0.00058}^{+0.00058}$ & $0.02048 \to 0.02416$
      \\[1mm]
       $\delta$
      & $(1.29_{-0.14}^{+0.20})\pi$ & $0.80 \pi \to 1.94 \pi$
      & $(1.53_{-0.16}^{+0.12})\pi$ & $1.08 \pi \to 1.91 \pi$
      \\[3mm]
       $\Delta m^2_{21}/(10^{-5}~{\rm eV}^2)$
      & $7.41_{-0.20}^{+0.21}$ & $6.82 \to 8.03$
      & $7.41_{-0.20}^{+0.21}$ & $6.82 \to 8.03$
      \\[3mm]
       $|\Delta m^2_{31}|/(10^{-3}~{\rm eV}^2)$
      & $2.507_{-0.027}^{+0.026}$ & $2.427 \to 2.590$
      & $2.412_{-0.025}^{+0.028}$ & $2.332 \to 2.496$
      \\[2mm]
      \hline\hline
    \end{tabular}
  \end{footnotesize}
  \caption{The best-fit values, 1$\sigma$ errors and 3$\sigma$ ranges of six neutrino
oscillation parameters extracted from a global analysis of the existing
neutrino oscillation data \cite{global1}. }
\end{table}

\section{Some basic formulas of leptogenesis}

In this section, we recapitulate some basic formulas of leptogenesis as a basis of our study.

In the classical high-scale thermal leptogenesis scenario, the amount of lepton-antilepton asymmetry produced crucially depends on the CP asymmetries between the right-handed neutrino decay processes $N^{}_I \to L^{}_\alpha + H$ (for $\alpha = e, \mu, \tau$) and their CP-conjugate processes $N^{}_I \to \overline{L}^{}_\alpha + \overline{H}$ (with $L^{}_\alpha$ and $H$ being respectively the lepton and Higgs doublets). In the case that the right-handed neutrino masses are hierarchical (i.e., $M^{}_1 \ll M^{}_2 \ll M^{}_3$), the flavor-specific CP asymmetries for the decays of $N^{}_I$ are given by
\begin{eqnarray}
&& \varepsilon^{}_{I \alpha} = \frac{1}{8\pi (M^\dagger_{\rm D}
M^{}_{\rm D})^{}_{II} v^2} \sum^{}_{J \neq I} \left\{ {\rm Im}\left[(M^*_{\rm D})^{}_{\alpha I} (M^{}_{\rm D})^{}_{\alpha J}
(M^\dagger_{\rm D} M^{}_{\rm D})^{}_{IJ}\right] {\cal F} \left( \frac{M^2_J}{M^2_I} \right) \right. \nonumber \\
&& \hspace{1.cm}
+ \left. {\rm Im}\left[(M^*_{\rm D})^{}_{\alpha I} (M^{}_{\rm D})^{}_{\alpha J} (M^\dagger_{\rm D} M^{}_{\rm D})^*_{IJ}\right] {\cal G}  \left( \frac{M^2_J}{M^2_I} \right) \right\} \; ,
\label{2.1}
\end{eqnarray}
with ${\cal F}(x) = \sqrt{x} \{(2-x)/(1-x)+ (1+x) \ln [x/(1+x)] \}$ and ${\cal G}(x) = 1/(1-x)$. The total CP asymmetry $\varepsilon^{}_I$ is obtained by summing the contributions from the three lepton flavours (i.e., $\varepsilon^{}_I = \sum^{}_\alpha \varepsilon^{}_{I \alpha}$).

As is known, according to the temperature ranges where leptogenesis takes place (approximately the right-handed neutrino mass scale), there are the following three distinct leptogenesis regimes  \cite{flavor1, flavor2}.
(1) Unflavored regime: in the temperature range above $10^{12}$ GeV where the charged-lepton Yukawa $y^{}_\alpha$ interactions are not in thermal equilibrium, the lepton flavors are indistinguishable. Consequently, leptogenesis proceeds entirely in the unflavoured regime, and three lepton flavors should be treated in a universal way. In the case that the right-handed neutrino masses are hierarchical,
the final baryon asymmetry mainly comes from the lightest right-handed neutrino $N^{}_1$, since its related processes will effectively washout the lepton asymmetries generated from the heavier right-handed neutrinos. And the final baryon asymmetry from it can be calculated according to
\begin{eqnarray}
Y^{}_{\rm B} = c Y^{}_{\rm L} = c r \varepsilon^{}_1 \kappa(\widetilde m^{}_1)  \;,
\label{2.2}
\end{eqnarray}
where $c \simeq -1/3$ is the conversion efficiency from the lepton asymmetry to the baryon asymmetry via the sphaleron processes, and $r \simeq 4 \times 10^{-3}$ measures the ratio of the equilibrium number density of $N^{}_I$ to the entropy density. Finally, $\kappa(\widetilde m^{}_1) \leq 1$ is the efficiency factor (i.e., the survival probability of the lepton asymmetry generated from the decays of $N^{}_1$) which takes account of the washout effects due to the inverse decays of $N^{}_1$ and various lepton-number-violating scattering processes. Its concrete value depends on the washout mass parameter
\begin{eqnarray}
\widetilde m^{}_I = \sum^{}_\alpha \widetilde m^{}_{I \alpha} = \sum^{}_\alpha  \frac{|(M^{}_{\rm D})^{}_{\alpha I}|^2}{M^{}_I} \;,
\label{2.3}
\end{eqnarray}
and can be numerically calculated by solving the relevant Boltzmann equations. In the strong washout regime which applies in most of realistic leptogenesis parameter space, the efficiency factor is roughly inversely proportional to the washout mass parameter.
(2) Two-flavor regime: in the temperature range $10^{9}-10^{12}$ GeV where the $y^{}_\tau$-related interactions are in thermal equilibrium while those of $e$ and $\mu$ are not, the $\tau$ flavor becomes distinguishable from the other two flavors. Therefore, there exist two distinct flavors: the $\tau$ flavor and a coherent superposition of the $e$ and $\mu$ flavors. In this regime, the final baryon asymmetry from $N^{}_1$ can be calculated according to
\begin{eqnarray}
Y^{}_{\rm B} = c(Y^{}_{{\rm L}\gamma} + Y^{}_{{\rm L}\tau} )
=  c r \left[ \varepsilon^{}_{1 \gamma} \kappa \left( \widetilde m^{}_{1 \gamma} \right) + \varepsilon^{}_{1 \tau} \kappa \left(\widetilde m^{}_{1 \tau} \right) \right]
 \;,
\label{2.4}
\end{eqnarray}
with $\varepsilon^{}_{I \gamma} = \varepsilon^{}_{I e} + \varepsilon^{}_{I \mu}$ and $\widetilde m^{}_{I \gamma} = \widetilde m^{}_{I e} + \widetilde m^{}_{I \mu}$. Here we use $Y^{}_{{\rm L}\alpha}$ to denote the lepton asymmetry stored in the $\alpha$ flavor. (3) Three-flavor regime: in the temperature range below $10^{9}$ GeV where the $y^{}_\mu$-related interactions are also in thermal equilibrium, all the flavors are distinguishable, thus each flavor should be treated separately. In this regime, the final baryon asymmetry from $N^{}_1$ can be calculated according to
\begin{eqnarray}
Y^{}_{\rm B} = c(Y^{}_{{\rm L}e} + Y^{}_{{\rm L}\mu} + Y^{}_{{\rm L}\tau} ) = c r \left[ \varepsilon^{}_{1 e} \kappa \left( \widetilde m^{}_{1 e} \right) + \varepsilon^{}_{1 \mu} \kappa \left( \widetilde m^{}_{1 \mu} \right) + \varepsilon^{}_{1 \tau} \kappa \left( \widetilde m^{}_{1\tau} \right) \right] \; .
\label{2.5}
\end{eqnarray}

It is known that in the case that the right-handed neutrino masses are hierarchial, in order to successfully reproduce the observed value of $Y^{}_{\rm B}$, there exists a lower bound about $10^9$ GeV for the right-handed neutrino mass scale \cite{DI}, which is too high to be directly accessed by forseeable collider experiments. Fortunately, a unique possibility to have a significantly lower leptogenesis scale is provided by the resonant leptogenesis scenario, which is realized in the case that the right-handed neutrino masses are nearly degenerate \cite{resonant1, resonant2}. In this scenario, the CP asymmetries will get resonantly enhanced and consequently the observed value of $Y^{}_{\rm B}$ can be successfully reproduced even for TeV scale right-handed neutrinos (which have the potential to be directly accessed by running or upcoming collider experiments \cite{RHN}). To be specific, in the resonant leptogenesis regime the flavor-specific CP asymmetries for the decays of $N^{}_I$ are given by
\begin{eqnarray}
\varepsilon^{}_{I\alpha} = \frac{{\rm Im}\left\{ (M^*_{\rm D})^{}_{\alpha I} (M^{}_{\rm D})^{}_{\alpha J}
\left[ M^{}_J (M^\dagger_{\rm D} M^{}_{\rm D})^{}_{IJ} + M^{}_I (M^\dagger_{\rm D} M^{}_{\rm D})^{}_{JI} \right] \right\} }{8\pi  v^2 (M^\dagger_{\rm D} M^{}_{\rm D})^{}_{II}} \cdot \frac{M^{}_I \Delta M^2_{IJ}}{(\Delta M^2_{IJ})^2 + M^2_I \Gamma^2_J} \;,
\label{2.6}
\end{eqnarray}
where $\Delta M^2_{IJ} \equiv M^2_I - M^2_J$ has been defined and $\Gamma^{}_J= (M^\dagger_{\rm D} M^{}_{\rm D})^{}_{JJ} M^{}_J/(8\pi v^2)$ is the decay rate of $N^{}_J$ (for $J \neq I$).
For the low scale resonant leptogenesis scenario, leptogenesis works in the three-flavor regime, and the contributions of the nearly degenerate right-handed neutrinos to the final baryon asymmetry are on the same footing and should be taken into consideration altogether. Correspondingly, the final baryon asymmetry is given by
\begin{eqnarray}
Y^{}_{\rm B}  = c(Y^{}_{{\rm L}e} + Y^{}_{{\rm L}\mu} + Y^{}_{{\rm L}\tau} ) = c r \left[ \kappa \left( \sum^{}_I \widetilde m^{}_{I e} \right) \sum^{}_I \varepsilon^{}_{I e} + \kappa \left( \sum^{}_I \widetilde m^{}_{I \mu} \right) \sum^{}_I \varepsilon^{}_{I \mu} + \kappa \left( \sum^{}_I \widetilde m^{}_{I \tau} \right) \sum^{}_I \varepsilon^{}_{I \tau} \right] \;.
\label{2.7}
\end{eqnarray}
Note that, for each lepton flavor, the washout is described by the sum of the related washout mass parameter for each right-handed neutrino.

Noteworthy, the baryon asymmetry does not necessarily have to be solely produced during the decay (freeze-out) processes of right-handed neutrinos. For GeV scale right-handed neutrinos, the baryon asymmetry is produced during their production (freeze-in) processes \cite{ARS1, ARS2}. This scenario is referred to as leptogenesis via oscillations or ARS leptogenesis. In this scenario, the semi-classical Boltzmann equation approach to thermal leptogenesis is insufficient, thus it is necessary to solve a set of quantum kinetic equations for the density matrices of right-handed neutrinos.
Due to the large dimensionality of the parameter space and complicated dynamics, it is difficult to provide a general analytical formula for the baryon asymmetry as a function of the model parameters. In the relevant calculations that follow, we make use of the latest version of the ULYSSES Python package \cite{uly} to obtain our results.

\section{Motivation of our study}

In this section, we first explain why it was thought that under the $\mu$-$\tau$ reflection symmetry the leptogenesis mechanism can only work in the two-flavor regime, and then introduce the flavor non-universality of the conversion efficiencies from the flavored lepton asymmetries to the baryon asymmetry which may help us achieve the successful leptogenesis in low scale seesaw models furnished with the $\mu$-$\tau$ reflection symmetry.

In the basis of the right-handed neutrino mass matrix $M^{}_{\rm R}$ being diagonal as $D^{}_{\rm R} = {\rm diag}(M^{}_1, M^{}_2, M^{}_3)$, under the $\mu$-$\tau$ reflection symmetry, the Dirac neutrino mass matrix $M^{}_{\rm D}$ takes a form as
\begin{eqnarray}
M^{}_{\rm D}= \displaystyle \left( \begin{matrix}
 a &  b & c \cr
a^\prime &  b^\prime & c^\prime \cr
 a^{\prime *} &  b^{\prime *} & c^{\prime *}
\end{matrix} \right) \left( \begin{matrix}
\sqrt{\eta^{}_1}  &  & \cr
 & \sqrt{\eta^{}_2}  &  \cr
 &  & \sqrt{\eta^{}_3}
\end{matrix} \right)  \;,
\label{3.1}
\end{eqnarray}
with $a$, $b$ and $c$ ($a^\prime$, $b^\prime$ and $c^\prime$) being real (complex) parameters and $\eta^{}_I = \pm 1$. It is easy to verify that such a form of $M^{}_{\rm D}$ and $D^{}_{\rm R}$ do keep invariant with respect to the transformations of three left-handed neutrino fields as described in Eq.~(\ref{4}) in combination with the transformations of three right-handed neutrino fields as $N^{}_I \leftrightarrow \eta^{}_I N^c_I$. Considering that the global phase of $M^{}_{\rm D}$ is of no physical meaning, in the following we simply take $\eta^{}_3 =1$.

Then, let us consider the implications of such a form of $M^{}_{\rm D}$ for leptogenesis. (1) For the case that the right-handed neutrino masses are hierarchial, by substituting Eq.~(\ref{3.1}) into Eqs.~(\ref{2.1}, \ref{2.3}), one arrives at
\begin{eqnarray}
\varepsilon^{}_{Ie} =0 \;, \hspace{1cm}  \varepsilon^{}_{I \mu} =- \varepsilon^{}_{I \tau} \;, \hspace{1cm} \widetilde m^{}_{I \mu} = \widetilde m^{}_{I \tau} \;.
\label{3.2}
\end{eqnarray}
From Eqs.~(\ref{2.2}, \ref{2.5}) one can immediately see that the leptogenesis mechanism cannot work (i.e., $Y^{}_{\rm B} =0$) in the unflavored regime as a result of $\varepsilon^{}_{I} =0$, and in the three-flavor regime as a result of
 \begin{eqnarray}
Y^{}_{{\rm L}e}=0\;, \hspace{1cm}  Y^{}_{{\rm L}\mu}= - Y^{}_{{\rm L}\tau} \;.
\label{3.3}
\end{eqnarray}
But Eq.~(\ref{2.4}) shows that, provided that $\widetilde m^{}_{1 e } \neq 0$ (i.e., $\widetilde m^{}_{1 \gamma} \neq \widetilde m^{}_{1 \tau}$), the leptogenesis mechanism can work in the two-flavor regime. (2) For the resonant leptogenesis regime, by substituting Eq.~(\ref{3.1}) into Eq.~(\ref{2.6}), one still arrives at the relations in Eq.~(\ref{3.2}) for $\eta^{}_I \eta^{}_J =1$, but $\varepsilon^{}_{I \alpha} \to 0$ for $\eta^{}_I \eta^{}_J =-1$ [see the discussions below Eq.~(\ref{4.8})]. Consequently, as can be seen from Eq.~(\ref{2.7}), the leptogenesis mechanism still cannot work in the three-flavor regime. (3) For the ARS leptogenesis regime, we have numerically checked that the flavored lepton asymmetries also obey the relations in Eq.~(\ref{3.3}).

As explained in the above, the leptogenesis mechanism cannot work  within low scale seesaw models furnished with the $\mu$-$\tau$ reflection symmetry. In order for the leptogenesis mechanism to work within such models, one needs to break the exact cancellation between the contributions of the $\mu$ and $\tau$ flavors to the baryon asymmetry.
For this problem, we put forward that the flavor non-universality of the conversion efficiencies from the flavored lepton asymmetries to the baryon asymmetry may play a crucial role:
in the above formulas of leptogenesis we have used a flavor universal conversion efficiency (i.e., $c\simeq -1/3$) from the lepton asymmetry to the baryon asymmetry, but in fact the hierarchies among the charged-lepton Yukawa couplings will lead to slightly different conversion efficiencies from the flavored lepton asymmetries to the baryon asymmetry. To be specific, the complete expression for the relation between the baryon and lepton asymmetries takes a form as (see Refs.~\cite{sphaleron1}-\cite{sphaleron4} for more details)
\begin{eqnarray}
Y^{}_{\rm B} & = & -4 \frac{77 T^2+54v^2}{869 T^2+666 v^2} \sum^{}_\alpha Y^{}_{{\rm L}\alpha}  \nonumber \\
&  & -
\left(\frac{11 v^2}{2\pi^2 T^2}
\frac{47 T^2+36 v^2}{869 T^2+666 v^2}
+ \frac{1}{16\pi^2} \frac{1034 T^2+810 v^2}{869 T^2+666 v ^2} \right)
\sum^{}_\alpha y_\alpha^2 Y^{}_{{\rm L}\alpha}  \;,
\label{3.4}
\end{eqnarray}
where $T \simeq 135$ GeV is the decoupling temperature of the sphaleron process \cite{sphaleron5}, $v=174$ GeV the Higgs VEV, and $y^{}_\alpha$ the charged-lepton Yukawa couplings. One can see that in the first term the conversion efficiency (which has a value about $-1/3$) from the lepton asymmetries to the baryon asymmetry is flavor universal and it is just the factor that is commonly used in the conventional researches. On the other hand, in the second term the conversion efficiencies from the flavored lepton asymmetries to the baryon asymmetry are flavor dependent (controlled by $y^2_\alpha$). Although the conversion coefficients in the second term are highly suppressed by $y^2_\alpha$ (concretely, $y^2_e \simeq 8.3 \times 10^{-12}$, $y^2_\mu \simeq 3.7 \times 10^{-7}$ and $y^2_\tau \simeq 1.0 \times 10^{-4}$), they may play a key role in the scenario that the total lepton asymmetry is vanishing: $Y^{}_{\rm L} =\sum^{}_\alpha Y^{}_{{\rm L}\alpha}=0$ [which would render the first term in Eq.~(\ref{3.4}) to be vanishing].
Thanks to such an effect, in the scenario considered in this paper which just realizes $\sum^{}_\alpha Y^{}_{{\rm L}\alpha}=0$, a non-vanishing $Y^{}_{\rm B}$ is still possible:
\begin{eqnarray}
Y^{}_{\rm B} \simeq - 0.06 (y^2_e Y^{}_{{\rm L} e} + y^2_\mu Y^{}_{{\rm L} \mu} + y^2_\tau Y^{}_{{\rm L} \tau} ) \simeq - 0.06 y^2_\tau Y^{}_{{\rm L} \tau} \;.
\label{3.5}
\end{eqnarray}
This tells us that, when $Y^{}_{{\rm L} \tau}$ is sizable enough (i.e., $- 1.4 \times 10^{-5}$), the observed value of $Y^{}_{\rm B}$ can be successfully reproduced. In the following two sections, we will study if the requisite baryon asymmetry can be successfully reproduced from such a mechanism for the resonant and ARS leptogenesis regimes, respectively.

\section{Study for resonant leptogenesis}

In this section, for the resonant leptogenesis regime, we study if the requisite baryon asymmetry can be successfully reproduced by means of the flavor non-universality of the conversion efficiencies from the flavored lepton asymmetries to the baryon asymmetry as shown in Eq.~(\ref{3.4}). In this section, we  perform the study in the general seesaw model with three right-handed neutrinos (in fact, as we will see the results indicate that the successful leptogenesis cannot be achieved in the minimal seesaw model with two right-handed neutrinos).

To make sure that we are in the correct parameter space region that reproduces the observed light neutrino masses and mixing, we use the widely-adopted Casas-Ibarra parametrization for $M^{}_{\rm D}$ \cite{CI1, CI2}. With the help of Eqs.~(\ref{1}, \ref{2}), one can easily understand that the Casas-Ibarra parametrization of $M^{}_{\rm D}$ takes a form as
\begin{eqnarray}
M^{}_{\rm D} = {\rm i} U D^{1/2}_\nu O D^{1/2}_{\rm R} \; ,
\label{4.1}
\end{eqnarray}
with $\sqrt{D^{}_\nu} = {\rm diag}(\sqrt{m^{}_1}, \sqrt{m^{}_2}, \sqrt{m^{}_3})$ and $\sqrt{D^{}_{\rm R}} = {\rm diag}(\sqrt{M^{}_1}, \sqrt{M^{}_2}, \sqrt{M^{}_3})$. Here $O$ contains the extra degrees of freedom of $M^{}_{\rm D}$ relative to $M^{}_\nu$ and satisfies the orthogonal relation $O^{T} O=I$.

Before proceeding, let us first consider the implications of the $\mu$-$\tau$ reflection symmetry [i.e., the particular form of $M^{}_{\rm D}$ in Eq.~(\ref{3.1})] for $U$ and $O$ in Eq.~(\ref{4.1}). Above all, taking account of the predictions of the $\mu$-$\tau$ reflection symmetry for the neutrino mixing parameters as given in Eq.~(\ref{5}), it is direct to see that $U$ takes a form as
\begin{eqnarray}
U^{}_{\mu\tau}  = \frac{1}{\sqrt{2}}
 \left( \begin{matrix}
\sqrt{2} c^{}_{12} c^{}_{13} & \sqrt{2} s^{}_{12} c^{}_{13} & - \sqrt{2} {\rm i} s^{}_{13}  \cr
-s^{}_{12}- {\rm i}  c^{}_{12} s^{}_{13}
& c^{}_{12} - {\rm i}  s^{}_{12} s^{}_{13}   & c^{}_{13} \cr
s^{}_{12} - {\rm i}  c^{}_{12} s^{}_{13}
& -c^{}_{12} - {\rm i} s^{}_{12} s^{}_{13}  & c^{}_{13}
\end{matrix} \right) \left( \begin{matrix}
\sqrt{\eta^{}_\rho} &  & \cr
& \sqrt{\eta^{}_\sigma}  & \cr
&  & 1
\end{matrix} \right) \;,
\label{4.2}
\end{eqnarray}
with $\eta^{}_\rho=1$ or $-1$ for $\rho=0$ or $\pi/2$ (and similarly for $\eta^{}_\sigma$). Then, by making a comparison between Eq.~(\ref{3.1}) and Eq.~(\ref{4.1}), one can further derive the following correspondence relations between $(\eta^{}_1, \eta^{}_2)$ and $(\rho, \sigma)$ and $O$:
\begin{eqnarray}
&& (\eta^{}_1, \eta^{}_2)=(+1, +1) \hspace{1cm} \Longleftrightarrow \hspace{1cm} (\rho, \sigma) =\left(\frac{\pi}{2}, \frac{\pi}{2} \right) \;, \hspace{0.5cm}  O = O^{}_x O^{}_y O^{}_z  \; ; \nonumber \\
&& (\eta^{}_1, \eta^{}_2)=(-1, +1) \hspace{1cm} \Longleftrightarrow \hspace{1cm} (\rho, \sigma) =\left(0, \frac{\pi}{2} \right) \;, \hspace{0.5cm}  O = O^{}_x O^{\prime}_y O^{\prime}_z  \; ; \nonumber \\
&& (\eta^{}_1, \eta^{}_2)=(+1, -1) \hspace{1cm} \Longleftrightarrow \hspace{1cm} (\rho, \sigma) =\left(\frac{\pi}{2}, 0 \right) \;, \hspace{0.5cm}  O = O^{\prime}_x O^{}_y O^{\prime}_z  \; ; \nonumber \\
&& (\eta^{}_1, \eta^{}_2)=(-1, -1) \hspace{1cm} \Longleftrightarrow \hspace{1cm} (\rho, \sigma) =\left(0, 0 \right) \;, \hspace{0.5cm}  O = O^{\prime}_x O^{\prime}_y O^{}_z  \; ,
\label{4.3}
\end{eqnarray}
with
\begin{eqnarray}
&& O^{}_x  = \left( \begin{matrix}
1 & 0 & 0 \cr
0 & \cos x  & \sin x \cr
0 & -\sin x &  \cos x
\end{matrix} \right) \;, \hspace{1cm}
O^\prime_x  = \left( \begin{matrix}
1 & 0 & 0 \cr
0 & \cosh x  & {\rm i}  \sinh x \cr
0 & - {\rm i} \sinh x &  \cosh x
\end{matrix} \right) \;, \nonumber \\
&& O^{}_y = \left( \begin{matrix}
\cos y & 0 &  \sin y \cr
0 & 1  & 0 \cr
-\sin y & 0 &  \cos y
\end{matrix} \right) \;, \hspace{1cm}
O^\prime_y = \left( \begin{matrix}
\cosh y & 0 & {\rm i}  \sinh y \cr
0 & 1  & 0 \cr
-{\rm i} \sinh y & 0 &  \cosh y
\end{matrix} \right) \;, \nonumber \\
&& O^{}_z = \left( \begin{matrix}
\cos z &  \sin z & 0 \cr
-\sin z &  \cos z  & 0 \cr
0 & 0 & 1
\end{matrix} \right) \;, \hspace{1cm}
O^\prime_z = \left( \begin{matrix}
\cosh z & {\rm i}  \sinh z & 0 \cr
-{\rm i} \sinh z &  \cosh z  & 0 \cr
0 & 0 & 1
\end{matrix} \right) \;,
\label{4.4}
\end{eqnarray}
where $x$, $y$ and $z$ are real parameters. In the following, for simplicity and clarity, we will just consider the cases that only one of $x$, $y$ and $z$ is non-vanishing (for model realizations of such cases, see Refs.~\cite{rCP1}-\cite{rCP3}).

Let us first perform the study for the case of $O=O^{}_x$ [corresponding to $\eta^{}_2 = +1$ and $\sigma =\pi/2$, see Eq.~(\ref{4.3})]. In this case, $N^{}_1$ will decouple from leptogenesis and its role is to be responsible for the generation of $m^{}_1$. This point can be easily understood by noting that such a form of $O$ will lead us to have $\varepsilon^{}_{1 \alpha} = 0$ as a result of $(M^\dagger_{\rm D} M^{}_{\rm D})^{}_{1J} =0$ for $J\neq 1$ [see Eq.~(\ref{2.6})]. Hence one just needs to consider the contributions of nearly degenerate $N^{}_2$ and $N^{}_3$ to the baryon asymmetry, and the $\tau$-flavor CP asymmetries for their decays are explicitly given by
\begin{eqnarray}
&& \varepsilon^{}_{2 \tau} \simeq \frac{(m^{}_3 - m^{}_2)\sqrt {m^{}_2 m^{}_3} \sin 2x}{8\pi v^2 (m^{}_2 \cos^2 x + m^{}_3 \sin^2 x) } \cdot \frac{M^{2}_0 \Delta M^{}_{32}}{4{{\left( {\Delta M^{}_{32}} \right)}^2} + \Gamma^2_3} \cdot c^{}_{12} c^{}_{13}\;,
 \nonumber \\
&& \varepsilon^{}_{3 \tau} \simeq \frac{(m^{}_3 - m^{}_2)\sqrt {m^{}_2 m^{}_3} \sin 2x}{8\pi v^2 (m^{}_2 \sin^2 x + m^{}_3 \cos^2 x) } \cdot \frac{M^{2}_0 \Delta M^{}_{32}}{4{{\left( {\Delta M^{}_{32}} \right)}^2} + \Gamma^2_2} \cdot c^{}_{12} c^{}_{13}\;,
\label{4.5}
\end{eqnarray}
with $\Delta M^{}_{IJ} = M^{}_I - M^{}_J$ and $M^{}_0 \simeq M^{}_2 \simeq M^{}_3$.
In the following, we will consider both the possibilities of the right-handed neutrinos being completely nearly degenerate (i.e., three right-handed neutrinos have nearly equal masses) and partially nearly degenerate (i.e., only two right-handed neutrinos have nearly equal masses).
For the present case, there are the following two possible mass spectra of the right-handed neutrinos: $M^{}_1 < M^{}_2 \approx M^{}_3$ and $M^{}_1 \approx M^{}_2 \approx M^{}_3$. It should be noted that the baryon asymmetries from $N^{}_2$ and $N^{}_3$ are subject to the washout effects from $N^{}_1$-related interactions: (1) For the possibility of $M^{}_1 < M^{}_2 \approx M^{}_3$, the final baryon asymmetry is given by \cite{N1wash1}-\cite{N1wash3}
\begin{eqnarray}
Y^{}_{\rm B} \simeq - 0.06 y^2_\tau r (\varepsilon^{}_{2 \tau} + \varepsilon^{}_{3 \tau}) \kappa \left( \widetilde m^{}_{2\tau} +  \widetilde m^{}_{3\tau} \right)  e^{- \frac{3 K^{}_{1\tau}}{8\pi} } \;,
\label{4.6}
\end{eqnarray}
with $K^{}_{1\tau} \equiv \tilde m^{}_{1\tau}/m^{}_*$ where $m^{}_* \simeq 1.08 \times 10^{-3}$ eV is the so-called equilibrium neutrino mass. Figure~1(a) and (b) show the allowed values of $Y^{}_{\rm B}$ as functions of the lightest neutrino mass ($m^{}_{1}$ or $m^{}_3$) for the NO and IO case, respectively. In obtaining these results, we have taken the following parameter configurations (and similarly in the following): We have employed the data in Table~1 for the neutrino mixing angles $\theta^{}_{12}$ and $\theta^{}_{13}$ and neutrino mass-squared differences, and fixed $\theta^{}_{23}$ and $\delta$ to be $\pi/4$ and $-\pi/2$ as predicted by the $\mu$-$\tau$ reflection symmetry. For the nearly degenerate right-handed neutrinos $N^{}_2$ and $N^{}_3$, we have taken their common mass $M^{}_0=1$ TeV as a benchmark value and allowed their mass difference $\Delta M^{}_{32}$ to vary freely. Note that, although we have taken $M^{}_0=1$ TeV as a benchmark value, the allowed range of $Y^{}_{\rm B}$ will keep invariant for other values of $M^{}_0$ provided that $\Delta M^{}_{32}$ is allowed to vary freely. This can be easily seen from Eq.~(\ref{4.5}): in the expressions of $\varepsilon^{}_{I\tau}$, $\Delta M^{}_{32}$ and $M^{}_0$ only take effect in the form of $\Delta M^{}_{32}/M^2_0$ (note that $\Gamma^{}_I$ is proportional to $M^{2}_I$). Therefore, even when $M^{}_0$ and $\Delta M^{}_{32}$ change individually, the results of $Y^{}_{\rm B}$ will be the same provided that their combination $\Delta M^{}_{32}/M^2_0$ keeps invariant. Finally, we have allowed $x$ to vary in the range $0-2\pi$. The results show that in the NO case, the maximally allowed values of $Y^{}_{\rm B}$ are approximately half of the observed value, and exhibit a significant suppression for large values of $m^{}_1$ [this is because in the present case one has $\tilde m^{}_1 = m^{}_1$ so that the washout effects from the $N^{}_1$-related interactions exponentially grow with $m^{}_1$ as shown by Eq.~(\ref{4.6})]. In the IO case, the allowed values of $Y^{}_{\rm B}$ are suppressed more severely. This is attributed to the fact that $m^{}_1$ is larger in the IO case compared to the NO case, resulting in stronger washout effects from $N^{}_1$-related interactions. (2) For the possibility of $M^{}_1 \approx M^{}_2 \approx M^{}_3$, the final baryon asymmetry is given by
\begin{eqnarray}
Y^{}_{\rm B}  = - 0.06 y^2_\tau r (\varepsilon^{}_{2\tau} + \varepsilon^{}_{3\tau}) \kappa \left( \widetilde m^{}_{1\tau} +  \widetilde m^{}_{2\tau} +  \widetilde m^{}_{3\tau} \right) \;.
\label{4.7}
\end{eqnarray}
Figure~1(c) and (d) show the allowed values of $Y^{}_{\rm B}$ as functions of the lightest neutrino mass for the NO and IO case, respectively.
The results show that the maximally allowed values of $Y^{}_{\rm B}$ are approximately half of the observed value.

\begin{figure*}
\centering
\includegraphics[width=6.5in]{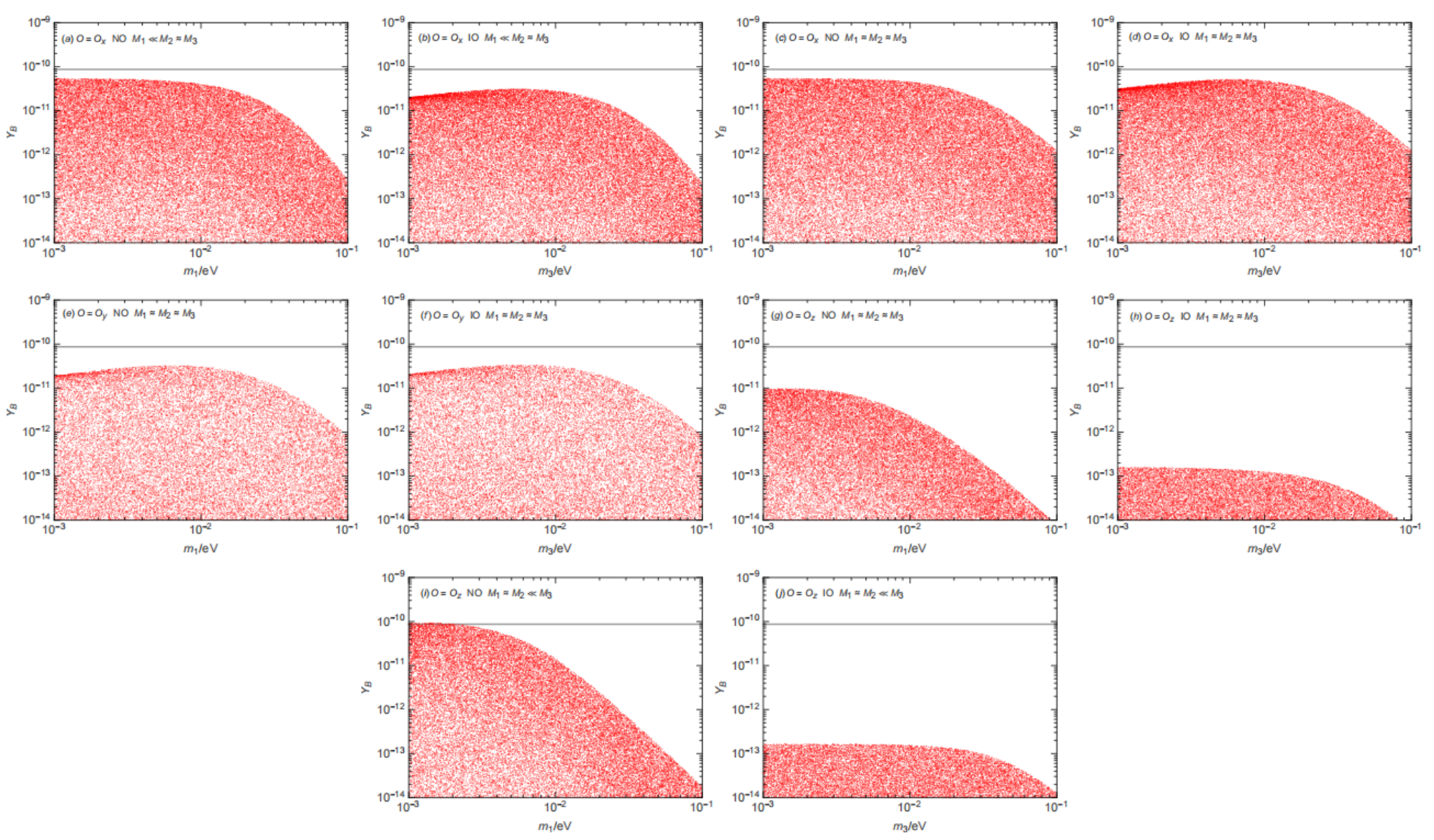}
\caption{ For the cases of $O^{} = O^{}_x, O^{}_y$ and $O^{}_z$ in Eq.~(\ref{4.4}) in combination with all the possible right-handed neutrino mass spectrum, the allowed values of $Y^{}_{\rm B}$ as functions of the lightest neutrino mass $m^{}_1$ and $m^{}_3$ in the NO and IO cases. The horizontal line stands for the observed value of $Y^{}_{\rm B}$. }
\label{fig1}
\end{figure*}

Then, let us consider the case of $O=O^\prime_x$ (corresponding to $\eta^{}_2 = -1$ and $\sigma =0$). In this case, $\varepsilon^{}_{2 \tau}$ and $\varepsilon^{}_{3 \tau}$ become
\begin{eqnarray}
&& \varepsilon^{}_{2\tau} = \frac{ \sqrt{m^{}_2 m^{}_3} (m^{}_2 + m^{}_3) \sinh 2 x }{16\pi  v^2 (m^{}_2 \cosh^2 x+ m^{}_3 \sinh^2 x)} \cdot \frac{M^{}_0 (\Delta M^{}_{32})^2}{4(\Delta M^{}_{32})^2 +  \Gamma^2_3}  \cdot c^{}_{12} c^{}_{13} \; , \nonumber \\
&& \varepsilon^{}_{3\tau} = \frac{ \sqrt{m^{}_2 m^{}_3} (m^{}_2 + m^{}_3)  \sinh 2 x }{16\pi  v^2 (m^{}_2 \cosh^2 x+ m^{}_3 \sinh^2 x)} \cdot \frac{M^{}_0 (\Delta M^{}_{32})^2 }{4(\Delta M^{}_{32})^2 +  \Gamma^2_2}   \cdot c^{}_{12} c^{}_{13} \; .
\label{4.8}
\end{eqnarray}
A comparison between Eq.~(\ref{4.8}) and Eq.~(\ref{4.5}) shows that $\varepsilon^{}_{I\tau}$ are additionally suppressed by $\Delta M^{}_{32}/M^{}_0$ (which should be very small as required by the resonance condition) in the present case compared to in the case of $O = O^{}_x$, making that the observed value of $Y^{}_{\rm B}$  definitely cannot be reached.
The results for the cases of $O = O^\prime_y$ and $O^\prime_z$ are similar, and we will not present them in the following.

In the case of $O=O^{}_y$ (corresponding to $\eta^{}_1 = +1$ and $\rho =\pi/2$), $N^{}_2$ will decouple from leptogenesis (i.e., $\varepsilon^{}_{2\alpha} =0$) and its role is to be responsible for the generation of $m^{}_2$. But the nearly degenerate right-handed neutrinos $N^{}_1$ and $N^{}_3$ may contribute to the final baryon asymmetry, and the $\tau$-flavor CP asymmetries for their decays are explicitly given by
\begin{eqnarray}
\varepsilon^{}_{1 \tau} \simeq -\frac{(m^{}_3 - m^{}_1)\sqrt {m^{}_1 m^{}_3} \sin 2y}{8\pi v^2 (m^{}_1 \cos^2 y + m^{}_3 \sin^2 y) } \cdot \frac{M^{2}_0 \Delta M^{}_{31}}{4{{\left( {\Delta M^{}_{31}} \right)}^2} + \Gamma^2_3} \cdot s^{}_{12} c^{}_{13}\;,
 \nonumber \\
\varepsilon^{}_{3 \tau} \simeq -\frac{(m^{}_3 - m^{}_1)\sqrt {m^{}_1 m^{}_3} \sin 2y}{8\pi v^2 (m^{}_1 \sin^2 y + m^{}_3 \cos^2 y) } \cdot \frac{M^{2}_0 \Delta M^{}_{31}}{4{{\left( {\Delta M^{}_{31}} \right)}^2} + \Gamma^2_1} \cdot s^{}_{12} c^{}_{13}\;.
\label{4.9}
\end{eqnarray}
with $M^{}_0 \simeq M^{}_1 \simeq M^{}_3$. In the present case, only the right-handed neutrino mass spectrum $M^{}_1 \approx M^{}_2 \approx M^{}_3$ is possible, and the final baryon asymmetry is correspondingly given by
\begin{eqnarray}
Y^{}_{\rm B}  = - 0.06 y^2_\tau r  (\varepsilon^{}_{1\tau} + \varepsilon^{}_{3\tau}) \kappa \left( \widetilde m^{}_{1\tau} +  \widetilde m^{}_{2\tau} +  \widetilde m^{}_{3\tau} \right) \;.
\label{4.10}
\end{eqnarray}
Figure~1(e) and (f) show the allowed values of $Y^{}_{\rm B}$ as functions of the lightest neutrino mass for the NO and IO case, respectively. The results show that the maximally allowed values of $Y^{}_{\rm B}$ are approximately one-third of the observed value.

In the case of $O = O^{}_z$ (corresponding to $\eta^{}_1 \eta^{}_2 = +1$ and $\rho -\sigma =0$), $N^{}_3$ will decouple from leptogenesis (i.e., $\varepsilon^{}_{3\alpha} =0$) and its role is to be responsible for the generation of $m^{}_3$. But the nearly degenerate right-handed neutrinos $N^{}_1$ and $N^{}_2$ may contribute to the final baryon asymmetry, and the $\tau$-flavor CP asymmetries for their decays are explicitly given by
\begin{eqnarray}
&& \varepsilon^{}_{1 \tau} \simeq \frac{(m^{}_2 - m^{}_1)\sqrt {m^{}_1 m^{}_2} \sin 2z}{8\pi v^2 (m^{}_1 \cos^2 z + m^{}_2 \sin^2 z) } \cdot \frac{M^{2}_0 \Delta M^{}_{21}}{4{{\left( {\Delta M^{}_{21}} \right)}^2} + \Gamma^2_2} \cdot s^{}_{13}\;,
 \nonumber \\
&& \varepsilon^{}_{2 \tau} \simeq \frac{(m^{}_2 - m^{}_1)\sqrt {m^{}_1 m^{}_2} \sin 2z}{8\pi v^2 (m^{}_1 \sin^2 z + m^{}_2 \cos^2 z) } \cdot \frac{M^{2}_0 \Delta M^{}_{21}}{4{{\left( {\Delta M^{}_{21}} \right)}^2} + \Gamma^2_1} \cdot s^{}_{13}\;.
\label{4.11}
\end{eqnarray}
with $M^{}_0 \simeq M^{}_1 \simeq M^{}_2$. In the present case, there are the following two possibilities for the right-handed neutrino mass spectrum: $M^{}_1 \approx M^{}_2 \approx M^{}_3$ and $M^{}_1 \approx M^{}_2 < M^{}_3$. (1) For the possibility of $M^{}_1 \approx M^{}_2 \approx M^{}_3$, the baryon asymmetries from $N^{}_1$ and $N^{}_2$ are subject to the washout effects from $N^{}_3$-related interactions, and the final baryon asymmetry is given by
\begin{eqnarray}
Y^{}_{\rm B}  = - 0.06 y^2_\tau r (\varepsilon^{}_{1\tau} + \varepsilon^{}_{2\tau}) \kappa \left( \widetilde m^{}_{1\tau} +  \widetilde m^{}_{2\tau} +  \widetilde m^{}_{3\tau} \right) \;.
\label{4.12}
\end{eqnarray}
Figure~1(i) and (j) show the allowed values of $Y^{}_{\rm B}$ as functions of the lightest neutrino mass for the NO and IO case, respectively. The results show that the maximally allowed values of $Y^{}_{\rm B}$ are smaller than the observed value, exhibiting a discrepancy of one order of magnitude in the NO case and a discrepancy of three orders of magnitude in the IO case.
(2) For the possibility of $M^{}_1 \approx M^{}_2 < M^{}_3$, the baryon asymmetries from $N^{}_1$ and $N^{}_2$ are not subject to the washout effects from $N^{}_3$-related interaction any more, and the final baryon asymmetry is given by
\begin{eqnarray}
Y^{}_{\rm B}  = - 0.06 y^2_\tau r (\varepsilon^{}_{1\tau} + \varepsilon^{}_{2\tau}) \kappa \left( \widetilde m^{}_{1\tau} +  \widetilde m^{}_{2\tau} \right)  \;.
\label{4.13}
\end{eqnarray}
Figure~1(g) and (h) show the allowed values of $Y^{}_{\rm B}$ as functions of the lightest neutrino mass for the NO and IO case, respectively. The results show that in the IO case the allowed values of $Y^{}_{\rm B}$ are three orders of magnitude smaller than the observed value. But in the NO case the observed value of $Y^{}_{\rm B}$ can be successfully reproduced. For this possibility, in Figure~2 we have further shown the parameter space of $\Delta M^{}_{21}/M^{}_0$ (i.e., the degeneracy level of $N^{}_1$ and $N^{}_2$) versus $m^{}_1$ for successful leptogenesis. One can see that in order to achieve a successful leptogenesis $m^{}_1$ should be within the range $0.7-2$ meV and $\Delta M^{}_{21}/M^{}_0$ should be within the range $(2-5) \times 10^{-15}$.
Note that these results are also obtained by taking $M^{}_0=1$ TeV as the benchmark value. But as mentioned below Eq.~(\ref{4.6}), for other values of $M^{}_0$, the results of $\Delta M^{}_{21}/M^{}_0$ can be obtained from those in Figure~2 by means of a simple rescaling law (e.g., will be lifted by 10 times for $M^{}_0=10$ TeV).
Before closing this section, we would like to point out that the above results indicate that in the minimal seesaw framework (with $m^{}_1=0$ or $m^{}_3=0$ in the NO or IO case) the observed value of $Y^{}_{\rm B}$ cannot be successfully reproduced in the resonant leptogenesis regime.

\begin{figure*}
\centering
\includegraphics[width=3.5in]{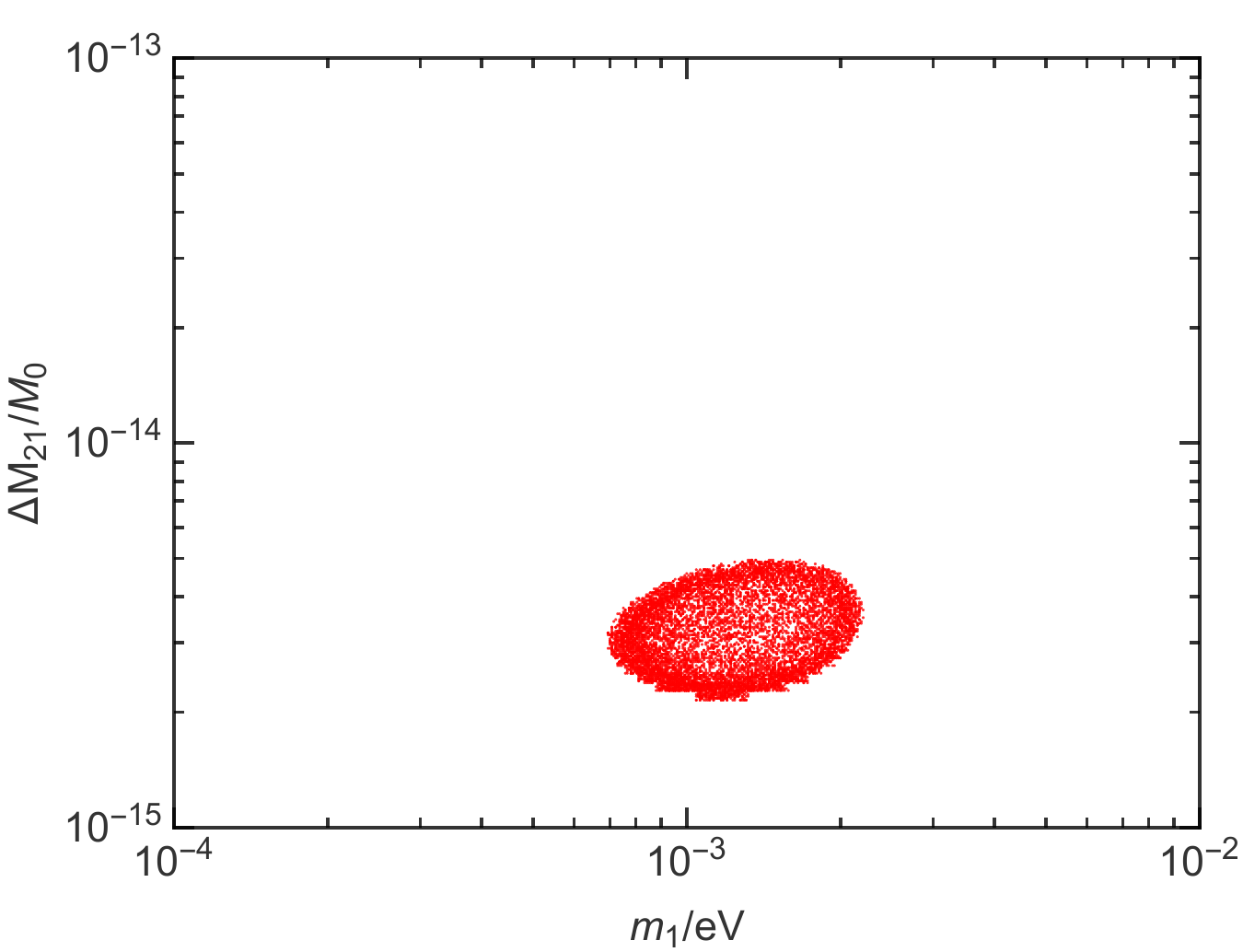}
\caption{ In the NO case, for the case of $O=O^{}_z$ in Eq.~(\ref{4.4}) and the right-handed neutrino mass spectrum $M^{}_1 \approx M^{}_2 < M^{}_3$, the parameter space of $\Delta M^{}_{21}/M^{}_0$ versus $m^{}_1$ for successful leptogenesis.  }
\label{fig1}
\end{figure*}

\section{Study for ARS leptogenesis}

In this section, for the ARS leptogenesis regime, we study if the requisite baryon asymmetry can be successfully reproduced by means of the flavor non-universality of the conversion efficiencies from the flavored lepton asymmetries to the baryon asymmetry as shown in Eq.~(\ref{3.4}). Among the models that work within the ARS leptogenesis regime, the most attractive and studied one is the so-called $\nu$MSM model \cite{ARS2, nuMSM2, nuMSM3}. In this model, the SM is extended with three right-handed neutrinos. But the lightest one of them serves as a keV-scale dark matter and is irrelevant to the generation of light neutrino masses, while the heavier two GeV-scale right-handed neutrinos are responsible for the generations of light neutrino masses as well as the baryon asymmetry through the ARS leptogenesis scenario. Effectively, this model belongs to the class of minimal seesaw model (with only two right-handed neutrinos relevant to the generation of light neutrino masses) which has attracted a lot of attention due to its simplicity \cite{MSS1}-\cite{MSS7}. For these considerations, in this section we will perform the study in the minimal seesaw model framework.

For the minimal seesaw model, in the basis of the right-handed neutrino mass matrix $M^{}_{\rm R}$ being diagonal as $D^{}_{\rm R} = {\rm diag}(M^{}_1, M^{}_2)$, under the $\mu$-$\tau$ reflection symmetry, the Dirac neutrino mass matrix $M^{}_{\rm D}$ takes a form as
\begin{eqnarray}
M^{}_{\rm D}= \displaystyle \left( \begin{matrix}
 a &  b \cr
a^\prime &  b^\prime  \cr
 a^{\prime *} &  b^{\prime *}
\end{matrix} \right) \left( \begin{matrix}
\sqrt{\eta}  &   \cr
 & 1
\end{matrix} \right)  \;,
\label{5.1}
\end{eqnarray}
with $a$ and $b$ ($a^\prime$ and $b^\prime$) being real (complex) parameters and $\eta = \pm 1$. It is known that the minimal seesaw model predicts the lightest neutrino to remain massless (i.e., $m^{}_1=0$ or $m^{}_3=0$ in the NO or IO case), and consequently only one or their combination of the Majorana CP phases is of physical meaning: $\sigma$ or $\rho-\sigma$ in the NO or IO case.
Then, taking account of the predictions of the $\mu$-$\tau$ reflection symmetry for the neutrino mixing matrix as shown in Eq.~(\ref{4.2}), a comparison between Eq.~(\ref{5.1}) and Eq.~(\ref{4.1}) revalues the following correspondence relations between $\eta$ and $\sigma$ and $O$:
\begin{eqnarray}
&& \eta =+1 \hspace{1cm} \Longleftrightarrow \hspace{1cm} \sigma =\frac{\pi}{2} \;, \hspace{0.5cm}  O= O^{}_1 \; ; \nonumber \\
&& \eta =-1 \hspace{1cm} \Longleftrightarrow \hspace{1cm} \sigma =0 \;, \hspace{0.5cm}   O=  O^{}_2  \;,
\label{5.2}
\end{eqnarray}
in the NO case, and
\begin{eqnarray}
&& \eta =+1 \hspace{1cm} \Longleftrightarrow \hspace{1cm} \rho- \sigma = 0 \;, \hspace{0.5cm}  O= O^{}_3 \; ; \nonumber \\
&& \eta =-1 \hspace{1cm} \Longleftrightarrow \hspace{1cm} \rho- \sigma =\frac{\pi}{2} \;, \hspace{0.5cm}  O= O^{}_4  \;,
\label{5.3}
\end{eqnarray}
in the IO case, with
\begin{eqnarray}
&&  O^{}_1=\left( \begin{matrix}
0 & 0 \cr
\cos \theta \ & \sin\theta \cr
- \sin \theta & \cos \theta
\end{matrix} \right) \; ;
\hspace{1cm}  O^{}_2 = \left( \begin{matrix}
0 & 0 \cr
\cosh \theta & {\rm i} \sinh \theta \cr
-{\rm i} \sinh \theta & \cosh \theta
\end{matrix} \right)  \;, \nonumber \\
&& O^{}_3=\left( \begin{matrix}
\cos \theta & \sin \theta \cr
-\sin \theta &   \cos \theta \cr
0 & 0
\end{matrix} \right) \; ;
\hspace{1cm}  O^{}_4 =\left( \begin{matrix}
\cosh \theta & {\rm i} \sinh \theta \cr
- {\rm i} \sinh \theta &  \cosh \theta  \cr
0 & 0
\end{matrix} \right)  \;,
\label{5.4}
\end{eqnarray}
where $\theta$ is a real parameter.

\begin{figure*}
\centering
\includegraphics[width=6.5in]{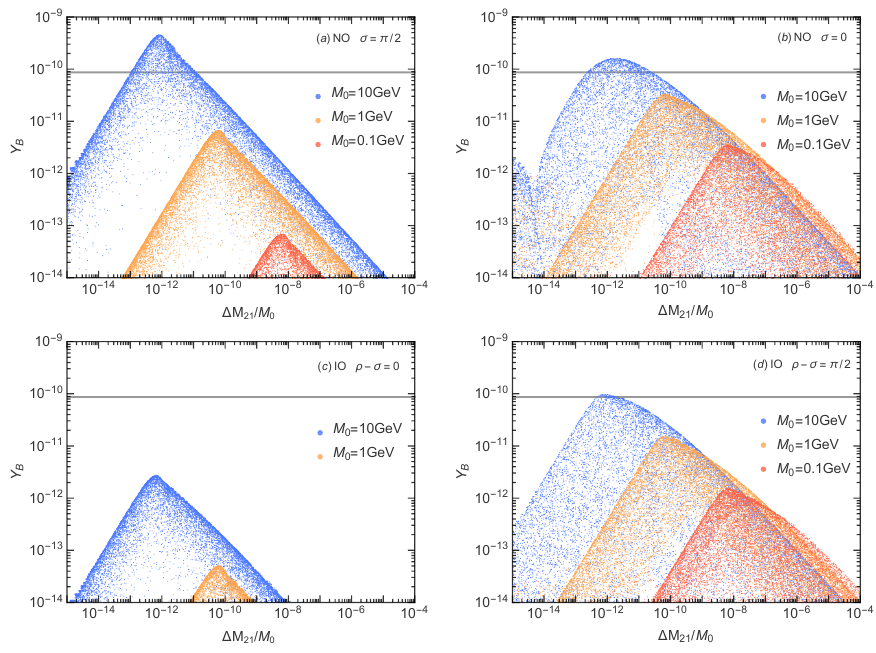}
\caption{ For the cases of $O =O^{}_1$, $O^{}_2$, $O^{}_3$ and $O^{}_4$ in Eqs.~(\ref{5.2}, \ref{5.3}), the allowed values of $Y^{}_{\rm B}$ as functions of $\Delta M^{}_{21}/ M^{}_0$ for some benchmark values of $M^{}_0$. The horizontal line stands for the observed value of $Y^{}_{\rm B}$. }
\label{fig3}
\end{figure*}

For the four cases listed in Eqs.~(\ref{5.2}, \ref{5.3}), Figure~3 shows the allowed values of $Y^{}_{\rm B}$ as functions of $\Delta M^{}_{21}/ M^{}_0$ (with $M^{}_0 \approx M^{}_1 \approx M^{}_2$ and $\Delta M^{}_{21} =  M^{}_2 -M^{}_1$) for some benchmark values of $M^{}_0$ in the range $0.1-100$ GeV. The reason why we have chosen to consider $M^{}_0$ in the range $0.1-100$ GeV is as follows: the lower bound 0.1 GeV for $M^{}_0$ comes from the constraint that the decays of right-handed neutrinos should not spoil the successful predictions of the Big Bang Nucleosynthesis \cite{RHN}; the upper bound 100 GeV for $M^{}_0$ comes from the consideration that one will eventually enter the resonant leptogenesis regime for $M^{}_0 \gtrsim 100$ GeV (and as mentioned above in the minimal seesaw framework the observed value of $Y^{}_{\rm B}$ cannot be successfully reproduced in the resonant leptogenesis regime). In obtaining the results, we have allowed $\theta$ to vary in the range $0-2\pi$ for the cases of $O=O^{}_1$ and $O^{}_3$. But for the cases of $O=O^{}_2$ and $O^{}_4$, we have allowed $\theta$ to vary in the range $[-3, 3]$ for the following consideration: large values of $x$ imply a strong fine tuning because they imply that neutrino masses are much lighter than the individual terms $(M^{}_{\rm D})^{2}_{\alpha I}/M^{}_I$ because of sign cancelations. Therefore, such choices tend to transfer the explanation of neutrino lightness from the seesaw mechanism to some other mechanism that has to explain the fine-tuned cancelations. A point of view held by Ref.~\cite{BB} is to consider the $O$ matrices to be "reasonable" if $|\sinh x|, |\sinh y|, |\sinh z| \lesssim 1$ (corresponding to $|x|, |y|, |z| \lesssim 1$), and "acceptable" if $|\sinh x|, |\sinh y|, |\sinh z| \lesssim 10$ (corresponding to $|x|, |y|, |z| \lesssim 3$). As shown in Figure~3(a) and (b), in the NO case both the cases of $O=O^{}_1$ and $O^{}_2$ can successfully reproduce the observed value of $Y^{}_{\rm B}$ for appropriate values of $M^{}_0$ and $\Delta M^{}_{21}$. For these two cases, in Figure~4(a) and (b) we have further shown the parameter space of $\Delta M^{}_{21}/M^{}_0$ versus $M^{}_0$ for successful leptogenesis. One can see that in order to achieve a successful leptogenesis $M^{}_0$ should be within the range $4-100$ GeV and $\Delta M^{}_{21}/M^{}_0$ should be within the range $10^{-15}-10^{-11}$. On the other hand, as shown in Figure~3(c) and (d), in the IO case the observed value of $Y^{}_{\rm B}$ can be successfully reproduced for the case of $O=O^{}_4$, but not for the case of $O=O^{}_3$. Nevertheless, for the case of $O=O^{}_4$, Figure~4(c) shows that the parameter space of $\Delta M^{}_{21}/M^{}_0$ versus $M^{}_0$ for successful leptogenesis is relatively narrow: $M^{}_0$ should be within the range $8-30$ GeV and $\Delta M^{}_{21}/M^{}_0$ should be within the range $10^{-13}- 3\times 10^{-12}$.

For right-handed neutrinos at the GeV scale, the quantum kinetic equations and reaction rates implemented in ULYSSES are expected to be valid. However, increasing the masses of right-handed neutrinos makes the lepton-number-violating terms more relevant, and the lepton-number-violating rates exhibit a pronounced temperature dependence near the electroweak crossover. Consequently, for right-handed neutrino masses in the range $M^{}_0 \geq 30$ GeV, the ULYSSES Python package incorporates the non-relativistic contributions to the lepton-number-violating rates as in Ref.~[59]. It is such a treatment that causes a discontinuity at exactly $M^{}_0 = 30$ GeV in Figure~4(b).

\begin{figure*}
\centering
\includegraphics[width=6.5in]{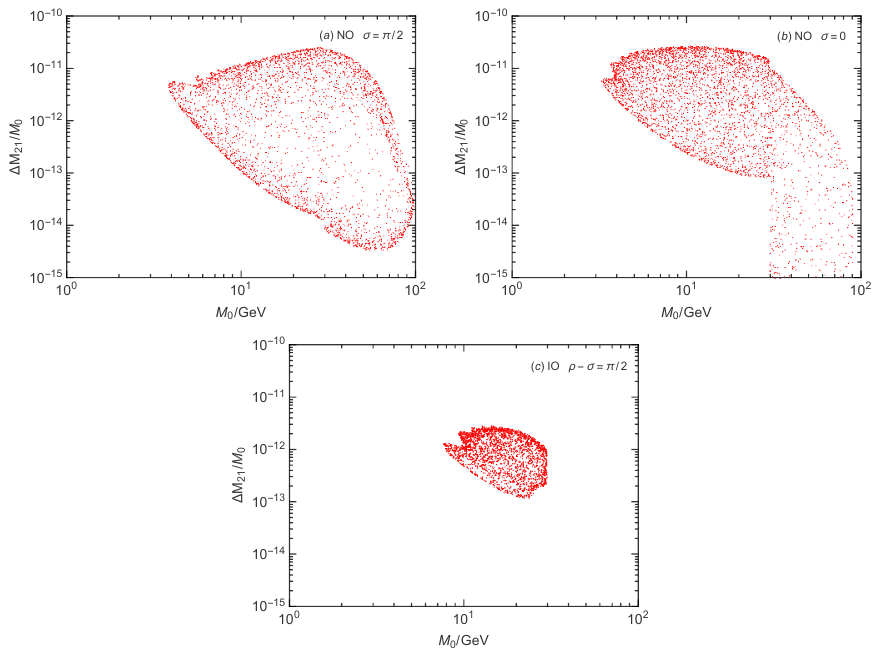}
\caption{ For the cases of $O =O^{}_1$, $O^{}_2$ and $O^{}_4$ in Eqs.~(\ref{5.2}, \ref{5.3}), the parameter space of $\Delta M^{}_{21}/M^{}_0$ versus $M^{}_0$ for successful leptogenesis. }
\label{fig3}
\end{figure*}

\section{Summary}

Due to the interesting predictions of the $\mu$-$\tau$ reflection symmetry for the neutrino mixing angles and CP phases as shown in Eq.~(\ref{5}), it is an attractive and widely studied candidate for the flavor symmetries in the neutrino sector. But it is known that, in the presence of this symmetry, the leptogenesis mechanism can only work in the two-flavor regime (which only holds for the right-handed neutrino masses in the range $10^9-10^{12}$ GeV), due to the exact cancellation between the contributions of the $\mu$ and $\tau$ flavors to the baryon asymmetry in the unflavored and three-flavor regimes.
This prohibits us to have a low scale seesaw model (which has the potential to be directly accessed by running or upcoming collider experiments) that can have the $\mu$-$\tau$ reflection symmetry and successful leptogenesis simultaneously.

In order for leptogenesis to work at low energies in the presence of the $\mu$-$\tau$ reflection symmetry, it is necessary to disturb the exact cancellation between the contributions of the $\mu$ and $\tau$ flavors to the baryon asymmetry. This can be achieved by taking account of the flavor non-universality of the conversion efficiencies from the flavored lepton asymmetries to the baryon asymmetry via the sphaleron process [see Eq.~(\ref{3.4})].
In this paper, for low scale seesaw models furnished with the $\mu$-$\tau$ reflection symmetry, we have demonstrated that the successful leptogenesis can be achieved in this way. We have performed the study in both the resonant leptogenesis regime (which is applicable for TeV scale right-handed neutrinos) and the ARS leptogenesis regime (which is applicable for GeV scale right-handed neutrinos).

For the resonant leptogenesis regime, we have performed the study in the general seesaw model with three right-handed neutrinos (in fact, the results indicate that the successful leptogenesis cannot be achieved in the minimal seesaw model with two right-handed neutrinos). In order to facilitate our study, we have first given the implications of the particular form of $M^{}_{\rm D}$ in Eq.~(\ref{3.1}) for $U$ and $O$ in the Casas-Ibarra parametrization in Eq.~(\ref{4.1}). The results are given in Eqs.~(\ref{4.2}-\ref{4.4}). For simplicity and clarity, we have just considered the cases that only one of $x$, $y$ and $z$ in Eq.~(\ref{4.4}) is non-vanishing. And we have considered both the possibilities of the right-handed neutrinos being completely nearly degenerate (i.e., three right-handed neutrinos have nearly equal masses) and partially nearly degenerate (i.e., only two right-handed neutrinos have nearly equal masses). The results show that only in the case of $O^{}_z$ in combination with the NO case of light neutrino masses and partially nearly degenerate right-handed neutrinos $M^{}_1 \approx M^{}_2 < M^{}_3$ can the observed value of $Y^{}_{\rm B}$ be successfully reproduced [see Figure~1(i)]. Figure~2 has further shown the parameter space for a successful leptogenesis: in order to achieve a successful leptogenesis $m^{}_1$ should be within the range $0.7-2$ meV.

For the ARS leptogenesis regime, we have performed the study in the minimal seesaw model, motivated by its simplicity and a potential embedding within the famous $\nu$MSM framework. The Casas-Ibarra parametrization results for the minimal seesaw model are given in Eqs.~(\ref{5.2}-\ref{5.4}).
The results show that in the NO case, both the cases of $O=O^{}_1$ and $O^{}_2$ can successfully reproduce the observed value of $Y^{}_{\rm B}$ for $M^{}_0$ within the range $4-100$ GeV and $\Delta M^{}_{21}$ within the range $10^{-15}-10^{-11}$. On the other hand, in the IO case the observed value of $Y^{}_{\rm B}$ can be successfully reproduced in the case of $O=O^{}_4$ for $M^{}_0$ within the range $8-30$ GeV and $\Delta M^{}_{21}/M^{}_0$ within the range $10^{-13}- 3\times 10^{-12}$, but not in the case of $O=O^{}_3$.

\vspace{0.5cm}

\underline{Acknowledgments} \vspace{0.2cm}

This work was supported in part by the National Natural Science Foundation of China under grant NO.~12475112, LiaoNing Revitalization Talents Program, and the Basic Research Business Fees for Universities in Liaoning Province under grant NO.~LJ212410165050.


\begin{thebibliography}{99}

\bibitem{xing} Z. Z. Xing, Phys. Rep. {\bf 854}, 1 (2020).

\bibitem{seesaw1} P. Minkowski, Phys. Lett. B {\bf 67}, 421 (1977).

\bibitem{seesaw2} M. Gell-Mann, P. Ramond and R. Slansky, in Supergravity, edited by P. van Nieuwenhuizen and D. Freedman, (North-Holland, 1979), p. 315.

\bibitem{seesaw3}  T. Yanagida, in Proceedings of the Workshop on the Unified Theory and the Baryon Number in the Universe, edited by O. Sawada and A. Sugamoto (KEK Report No. 79-18, Tsukuba, 1979), p. 95.

\bibitem{seesaw4} R. N. Mohapatra and G. Senjanovic, Phys. Rev. Lett. {\bf 44}, 912 (1980).

\bibitem{seesaw5} J. Schechter and J. W. F. Valle, Phys. Rev. D {\bf22}, 2227 (1980).

\bibitem{pmns1} B. Pontecorvo, Sov. Phys. JETP {\bf 6}, 429 (1957) [Zh. Eksp. Teor. Fiz. {\bf 33}, 549 (1957)].

\bibitem{pmns2} Z. Maki, M. Nakagawa and S. Sakata, Prog. Theor. Phys. {\bf 28}, 870 (1962).

\bibitem{global1} I. Esteban, M. C. Gonzalez-Garcia, M. Maltoni, T. Schwetz and A. Zhou, JHEP {\bf 09}, 178 (2020). NuFIT 5.2 (2022), www.nu-fit.org.

\bibitem{global2} F. Capozzi, E. D. Valentino, E. Lisi, A. Marrone, A. Melchiorri and A. Palazzo, Phys. Rev. D {\bf 104}, 083031 (2021).

\bibitem{global3} P. F. de Salas, D. V. Forero, S. Gariazzo, P. Martinez-Mirave, O. Mena, M. Tortola and J. W. F. Valle, JHEP {\bf 02}, 071 (2021).

\bibitem{T2K} K. Abe {\it et al.} (T2K Collaboration), Nature {\bf 580}, 339 (2020).

\bibitem{FS1} S. F. King and C. Luhn, Rept. Prog. Phys. {\bf 76}, 056201 (2013).

\bibitem{FS2} F. Feruglio and A. Romanino, Rev. Mod. Phys. {\bf 93}, 015007 (2021).

\bibitem{FS3} G. J. Ding, and S. F. King, arXiv:2311.09282.

\bibitem{FS4} G. J. Ding and J. W. F. Valle, arXiv:2402.16963.

\bibitem{mu-tauR1} P. H. Harrison and W. G. Scott, Phys. Lett. B {\bf 547}, 219 (2002).

\bibitem{mu-tauR2} Z. Z. Xing and Z. H. Zhao, Rept. Prog. Phys. {\bf 79}, 076201 (2016).

\bibitem{mu-tauR3} Z. Z. Xing, Rept. Prog. Phys. {\bf 86}, 076201 (2023).

\bibitem{leptogenesis} M. Fukugita and T. Yanagida, Phys. Lett. B {\bf 174}, 45 (1986).

\bibitem{Lreview1} W. Buchmuller, R. D. Peccei and T. Yanagida, Ann. Rev. Nucl. Part. Sci. {\bf 55}, 311 (2005).

\bibitem{Lreview2} W. Buchmuller, P. Di Bari and M. Plumacher, Annals Phys. {\bf 315}, 305 (2005).

\bibitem{Lreview3} S. Davidson, E. Nardi and Y. Nir, Phys. Rept. {\bf 466}, 105 (2008).

\bibitem{Lreview4} D. Bodeker and W. Buchmuller, Rev. Mod. Phys. {\bf 93}, 035004 (2021).

\bibitem{planck} P. A. R. Ade {\it et al.} (Planck Collaboration), Astron. Astrophys. A {\bf16}, 571 (2014).

\bibitem{DI} S. Davidson and A. Ibarra, Phys. Lett. B {\bf 535}, 25 (2002).

\bibitem{resonant1} A. Pilaftsis, Phys. Rev. D {\bf 56}, 5431 (1997).

\bibitem{resonant2} A. Pilaftsis and T. E. J. Underwood, Nucl. Phys. B {\bf 692}, 303 (2004).

\bibitem{ARS1} E. K. Akhmedov, V. A. Rubakov and A. Y. Smirnov, Phys. Rev. Lett. {\bf 81}, 1359 (1998).

\bibitem{ARS2} T. Asaka and M. Shaposhnikov, Phys. Lett. B {\bf 620}, 17 (2005).

\bibitem{RHN} A. M. Abdullahi {\it et al.}, J. Phys. G {\bf 50}, 020501 (2023).

\bibitem{MN} R. N. Mohapatra and C. C. Nishi, JHEP {\bf 08}, 092 (2015).

\bibitem{sphaleron1} V. A. Kuzmin, V. A. Rubakov and M. E. Shaposhnikov, Phys. Lett. B {\bf 191}, 171 (1987).

\bibitem{sphaleron2} S. Y. Khlebnikov and M. E. Shaposhnikov, Nucl. Phys. B {\bf 308}, 885 (1988).

\bibitem{sphaleron3} M. Laine and M. E. Shaposhnikov, Phys. Rev. D {\bf 61}, 117302 (2000).

\bibitem{sphaleron4} K. Mukaida, K. Schmitz and M. Yamada, Phys. Rev. Lett. {\bf 129}, 011803 (2022).

\bibitem{flavor1} A. Abada, S. Davidson, F. X. Josse-Michaux, M. Losada and A. Riotto, JCAP {\bf0604}, 004 (2006).

\bibitem{flavor2} E. Nardi, Y. Nir, E. Roulet and J. Racker, JHEP {\bf0601}, 164 (2006).

\bibitem{uly} A. Granelli, C. Leslie, Y. F. Perez-Gonzalez, H. Schulz, B. Shuve, J. Turner and R. Walker, Comput. Phys. Commun. {\bf 291}, 108834 (2023).

\bibitem{sphaleron5} M. D'Onofrio, K. Rummukainen and A. Tranberg, Phys. Rev. Lett. {\bf 113}, 141602 (2014).

\bibitem{CI1} J. A. Casas and A. Ibarra, Nucl. Phys. B {\bf 618}, 171 (2001).

\bibitem{CI2} A. Ibarra and G. G. Ross, Phys. Lett. B {\bf 591}, 285 (2004).

\bibitem{rCP1} P. Chen, G. J. Ding and S. F. King, JHEP {\bf 03}, 206 (2016).

\bibitem{rCP2} C. Hagedorn and E. Molinaro, Nucl. Phys. B {\bf 919}, 404 (2017).

\bibitem{rCP3} Z. H. Zhao, Eur. Phys. J. C {\bf 82}, 436 (2022).

\bibitem{N1wash1} S. Blanchet and P. Di Bari, JCAP {\bf 0606}, 023 (2006).

\bibitem{N1wash2} S. Blanchet and P. Di Bari, JCAP {\bf 03}, 018 (2007).

\bibitem{N1wash3} O. Vives, Phys. Rev. D {\bf 73}, 073006 (2006).

\bibitem{nuMSM2} T. Asaka, S. Blanchet and M. Shaposhnikov, Phys. Lett. B {\bf 631}, 151 (2005).

\bibitem{nuMSM3} M. Shaposhnikov, JHEP {\bf 08}, 008 (2008).

\bibitem{MSS1} A. Yu. Smirnov, Phys. Rev. D {\bf48}, 3264 (1993).

\bibitem{MSS2} S. F. King, Nucl. Phys. B {\bf576}, 85 (2000).

\bibitem{MSS3} S. F. King, JHEP {\bf0209}, 011 (2002).

\bibitem{MSS4} P. H. Frampton, S. L. Glashow and T. Yanagida, Phys. Lett. B {\bf 548}, 119 (2002).

\bibitem{MSS5} T. Endoh, S. Kaneko, S. K. Kang, T. Morozumi and M. Tanimoto, Phys. Rev. Lett. {\bf 89}, 231601 (2002).

\bibitem{MSS6} V. Barger, D. A. Dicus, H. J. He and T. J. Li, Phys. Lett. B {\bf 583}, 173 (2004).

\bibitem{MSS7} Z. Z. Xing and Z. H. Zhao, Rept. Prog. Phys. {\bf 84}, 066201 (2021).

\bibitem{BB} S. Blanchet and P. Di Bari, Nucl. Phys. B {\bf 807}, 155 (2009).

\bibitem{non-r} P. Hernandez, J. Lopez-Pavon, N. Rius and S. Sandner, JHEP {\bf 12}, 012 (2022).

\end{thebibliography}
\end{document}